\documentclass[12pt,twoside,a4paper,fleqn]{article}
\usepackage[margin=0.5in,footskip=0.5in]{geometry}

\usepackage{epsfig}
\usepackage{graphicx}
\usepackage{amssymb}
\usepackage{mathrsfs}
\usepackage{dcolumn}
\usepackage{multirow}

\newcommand{\pr}{\partial}
\newcommand{\rta}{\rightarrow}

\newcommand{\ep}{\epsilon}

\newcommand{\p}{\prime}
\newcommand{\om}{\omega}
\newcommand{\ra}{\rangle}
\newcommand{\la}{\langle}

\newcommand{\mcl}{\mathcal{L}}
\newcommand{\mcP}{\mathcal{P}}
\newcommand{\mcQ}{\mathcal{Q}}

\newcommand{\beq}{\begin{equation}}
\newcommand{\eeq}{\end{equation}}
\newcommand{\taud}{\tau_{Drude}}
\newcommand{\tbfk}{\textbf{k}}

\newcommand{\tbfq}{\textbf{q}}
\newcommand{\tbfp}{\textbf{p}}
\newcommand{\tbfkp}{{\textbf{k}^\p}}

\newcommand{\tbfpp}{{\textbf{p}^\p}}

\newcommand{\intmipi}{\int_{-\infty}^{+\infty}}

\newcommand{\ball}{\begin{align}}
\newcommand{\eall}{\end{align}}

\newcommand{\beqar}{\begin{eqnarray}}
\newcommand{\eeqar}{\end{eqnarray}}

\newcommand{\smnw}{\sigma_{\mu\nu}(\om)}
\newcommand{\unit}{1\!\!1}
\newcommand{\ketjm}{| J_\mu\ra}
\newcommand{\dg}{\dagger}

\newcommand{\smnz}{\sigma_{\mu\nu}(z)}

\newcommand{\ben}{\begin{enumerate}}
\newcommand{\een}{\end{enumerate}}
\makeatletter
\newcommand*{\rom}[1]{\expandafter\@slowromancap\romannumeral #1@}
\makeatother

\begin{document}
\title{The Memory Function Formalism: A Review}
\author{ Komal Kumari$^1$, Navinder Singh$^2$ \footnote{$^1$Department~of~Physics,~Himachal~Pradesh~University,~Shimla,~India, PIN:171005.~Email:~sharmakomal611@gmail.com }
\footnote{ $^2$Physical~Research~Laboratory,~Ahmedabad,~India, PIN:380009.~Email:~navinder.phy@gmail.com;~navinder@prl.res.in}}
\maketitle
\begin{abstract}
An introduction to the Zwanzig-Mori-G\"{o}tze-W\"{o}lfle memory function formalism (or generalized Drude formalism) is presented. This formalism is used extensively in analyzing the experimentally obtained optical conductivity of strongly correlated systems like cuprates and Iron based superconductors etc.    For a broader perspective both the generalised Langevin equation approach and the projection operator approach for the memory function formalism are given. The G\"{o}tze-W\"{o}lfle perturbative expansion of memory function is presented and its application to the computation of the dynamical conductivity of metals is also reviewd. This review of the formalism contains all the mathematical details for pedagogical purposes. 
\end{abstract}

``The first processes, therefore, in the effectual studies of the sciences, must be ones of simplification and reduction of the results of previous investigations to a form in which the mind can grasp them.''

\hspace{15cm} --J.C. Maxwell
\begin{figure}[h!]
{\huge{The pioneers of the memory function formalism}}
\begin{tabular}{ccc}
\includegraphics[height = 5cm, width =5cm]{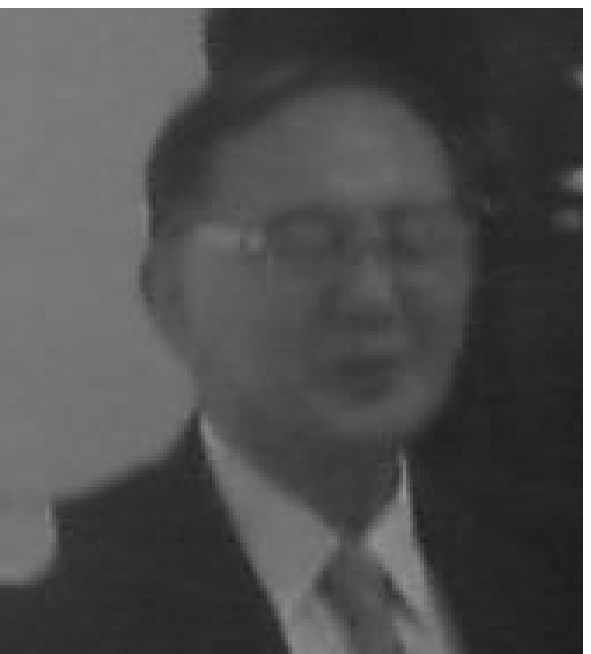}&~~
\includegraphics[height = 5cm, width =5cm]{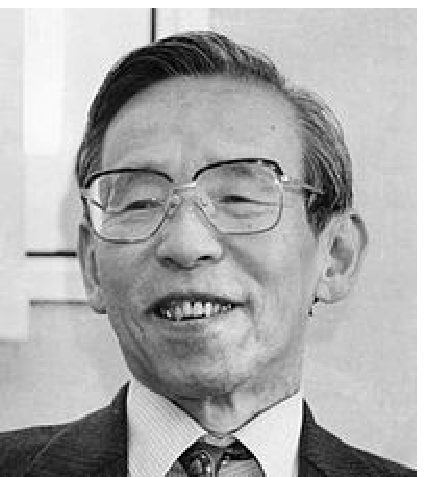}&~~
\includegraphics[height = 5cm, width =5cm]{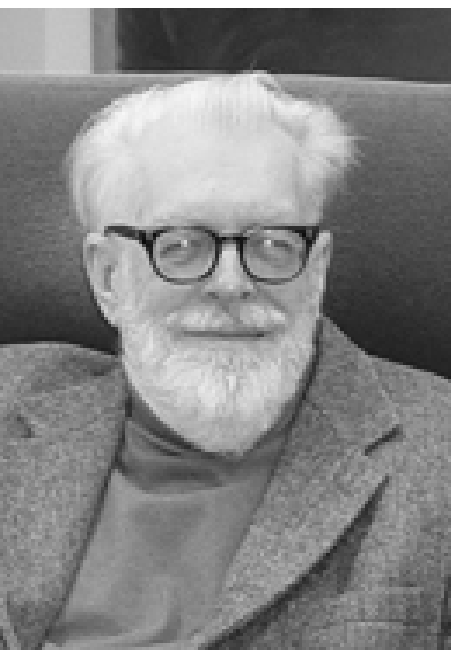}\\
(a)&
(b)&
(c)\\
\end{tabular}
\caption{(a) Hazime Mori (1926--) (b) Ryogo Kubo (1920--1995)  (c) Robert Zwanzig (1928-2014)}
\label{champs}
\end{figure}

\section{Introduction}
 A wider picture of the memory function formalism is presented. After sketching the standard derivation of the Drude formula of electrical conductivity from momentum relaxation equation, we introduce the Langevin equation for electrical conduction in metals. A derivation of the Drude formula from the Langevin equation is presented. By pointing out a fundamental problem with the Langevin equation, the problem of Drude formula is highlighted. Then generalised Langevin equation is introduced which rectifies the problem of the Langevin equation by incorporating the time dependent friction coefficient. Then generalised Drude formula is obtained from generalised  Langevin equation in which constant Drude scattering rate is replaced by a frequency dependent scattering rate, which is populary known as the memory function.\\
Then an alternative route to memory function formalism or generalised Drude formalism is sketched based on the Zwanzig-Mori projection opreator method. A formal expression for the memory function in terms of the projection operator is given. We then introduce the G\"{o}tze-W\"{o}lfle perturbation expansion of the memory function and review their formalism. Special cases of electron-impurity and electron-phonon interaction are revised in detail. We come back to the question of under what conditions frequency dependent Drude scattering rate can be taken as frequency independent parameter in standard Drude formula, and discuss the relevant conditions. Our overview provides a wider picture of this field of research, and nothing new is presented here. An attempt is made to put this beautiful field of research in a nutshell. We hope that this review will be useful for pedagogical purposes.

\section{The Drude Model}
The traditional way to introduce AC conductivity of a metal is to use the Drude formula. In the Drude model electrons in a metal are treated as classical particles (much like gas molecules) bumping with the stationary ion cores and constitute a random thermal motion. When an external electric field is applied, electrons are accelerated and gain momentum. The momentum gained from the field is dissipated due to the collisions with ion cores and they attain an average drift speed. Due to this finite drift speed an electric current is established in the sample. When external field is removed the drift speed vanishes and electrons go back to the thermally agitated motion. Thus under the action of applied field the equation of motion of electrons can be written as:
  
\begin{equation}
m\frac{d^{2}\textbf{r}}{dt^{2}}=-e\textbf{E}(t)-m 
\frac{1}{\tau}\frac{d\textbf{r}}{dt},
\label{motion}
\end{equation}
which is nothing but Newton's second law of motion including dissipation. The momentum of an electron is being degraded due to collision processes with the ion cores at the average rate $\frac{1}{\tau}$. Considering the electric field $\textbf{E}(t)$ variation  as  $\textbf{E}_0\cos(\om t)$ and solving the equation of motion for momentum leads to

\beq
\textbf{p}(t) = \underbrace{-e\frac{\tau}{1+ \tau^2 \om^2} \textbf{E}_0 
\cos(\om t)}_{In~phase~response~(dissipative)}  - \underbrace{e \frac{\om 
\tau^2}{1+ \tau^2\om^2}  \textbf{E}_0 \sin(\om 
t)}_{Out-of-phase~response~(reactive).}
\label{mom}
\eeq
One can observe that the time dependent momentum consitutes two responses: (1) the in-phase response leading to dissipation that varies as $\cos(\om t)$, and (2) out-of-phase response which has $\sin(\om t)$ dependence. From the above equation of momentum, the current density is given by: 
\beq
\textbf{j}(t) = - n e \frac{\textbf{p}(t)}{m} = \frac{n e^2 }{m}\bigg(\frac{\tau}{1+ \tau^2 \om^2} 
\textbf{E}_0 \cos(\om t) +\frac{\om \tau^2}{1+ \tau^2\om^2}  
\textbf{E}_0 \sin(\om t)\bigg). \label{mom1}
\eeq
The presence of current in the sample leads to dissipation of energy (Joule heating). The average heat dissipation is given as $\frac{\om}{2\pi}\int_0^{\frac{2\pi}{\om}} dt \textbf{j}(t).\textbf{E}(t)$. 
Clearly, this Joule heating is only due to the dissipative component, and it vanishes due to the reactive 
components (average of $\sin(\om t)\cos(\om t)$ over a period is zero). From equation (\ref{mom1}) the dissipative part of the conductivity ($\vec{J}=\sigma \vec{E})$ reduces to

\beq
\sigma_1(\om) = \sigma_{DC} \frac{1}{1+ \tau^2 \om^2}. \label{mom2}
\eeq

On the other hand  the reactive part of the conductivity becomes

\beq
\sigma_2(\om) = \sigma_{DC}\frac{\tau\om}{1+ \tau^2 \om^2}.
\eeq
In the following paragraph we will observe that Drude expression is valid when\\
1. Frequency ${\om}$ is very small as compared to the characterstics enegy scale in system (as discussed in section 6.2). \\
2. Scattering with ion cores is the dominant mechanism of momentum randomisation.\\
  Below we will derive the Drude formula using a different approach and we will observe that the scattering rate $\frac{1}{\tau}$ ceases to be a constant, rather it depends on frequency and temperature. 

\section{The Langevin equation}

In the present section we introduce an important subject of non-equilibrium statistical mechanics i.e., the Langevin equation, which provides the basis for the memory function(MF) formalism. Our purpose is to write ac conductivity (in a general case) in a form resembling the Drude formula derived from the Langevin equation. Such a  representation can be obtained using the MF formalism which will be our subject in the next sections. To understand memory function formalism we have to first consider a phenomenological approach. Let us start with the simple Langevin equation. 
The Langevin equation describe the random motion of a Brownian particle. The Brownian particle can be a pollen grain in some fluid. The origin of the random motion of Brownian particles is due to the irregular bombardment of the particle by molecules of fluid.  
It can also be viewed from the density fluctuations of  fluid at a spatial length scale of the Brownian particle size. 
The molecules of fluid also provide a drag force to Brownian particle along with the continuous random force. Thus, net force on the Brownian particles can be divided into two components (1) systematic drag force, and (2) random force. With this, equation of motion can be written as

\beq
m \dot{u}(t) = \underbrace{- m\gamma u(t)}_{systematic ~~part} + \underbrace{R(t)}_{random~~part.}
\label{lan1}
\eeq     
This is called the Langevin equation. This  equation holds  great significance not only to describe the motion of a Brownian particle, but its wide regime of applicability is  beautifully expressed in \cite{toda}
``Brownian motion is not merely random motion of a very fine particle; in general it is random motion  of a physical quantity to be observed in a macrosystem".
\vspace{5mm}

Let us apply the Langevin equation to electrons in a metal which is biased by an external ac field. The presence of impurity potentials and phonons\footnote{ The Johnson-Nyquist noise is a direct manifestation of this random motion.} force electrons to move randomly in the metal. On applying external ac field ($E_0 \cos(\om_0 t)$) the Langevin equation for an electron changes to
\beq
m \dot{u}(t) = - m \gamma u(t) + R(t) + E_0 \cos(\om_0 t).
\label{lan2}
\eeq
Here $m$ is the mass of an electron and $\gamma$ is the friction coefficient. Write the above equation in the Laplace domain with $u(s) = \int_0^\infty dt u(t) e^{- s t}$

\beq
- m u(0) + s m u(s) = -m \gamma u(s) + R(s) + \frac{E_0}{2} \bigg(\frac{1}{ s - i \om_0} + \frac{1}{s+i\om_0}\bigg).\label{lan(3)}
\eeq
On simplify it and performing the ensemble average we obtain
\beq
\la u(s)\ra = \frac{\la u(0)\ra}{s+\gamma} + \frac{E_0}{2m (s+\gamma)}\bigg(\frac{1}{ s - i \om_0} + \frac{1}{s+i\om_0}\bigg).\label{lan(4)}
\eeq
The ensemble average of the random force $\la R(t) \ra = 0$. The above equation in the long time limit $t>>\frac{1}{\gamma}$ can be written in the form
\beq
\la u(t)\ra = Re\{ \mu(\om_0) E_0 e^{i \om_0 t}\},\label{lan(5)}
\eeq
where
\beq
\mu(\om) = \frac{1}{m}\frac{1}{i\om +\gamma} \label{lan(6)}
\eeq
is the expression for the electronic mobility. In the electrical conduction problem the terminal speed along the $x-$direction (say) gained by an electron due to the applied electric field is $u_t^x = -\mu (e E_x)$ (terminal speed = mobility $\times$ force i.e., Fick's Law). The induced current density can be written in terms of mobility as $J_x = -n e u_t^x = n e^2 \mu E$. Accrding to Ohm's Law ${J_x}=\sigma_{x x} { E_x},$ thus $\sigma_{x x}=\sigma=n e^2\mu $. From equation (\ref{lan(6)}) of the mobility, the conductivity takes the form.
\beq
\sigma(\om) = \frac{ne^2}{m} \frac{1}{i\om +\gamma}. \label{lan(7)}
\eeq
The dc conductivity $\sigma(0) = n e^2/m\gamma$. Here ${\gamma}$ is to be identified with the Drude scattering rate $\frac{1}{\tau}$. This is the Drude formula that we derived in the previous section. From this derivation of the Drude formula an important insight can be gained. It will be shown in the following section that Langevin equation has a serious problem in a general setting. An understanding of the problem the Langevin equation will enable us to understand the problem of the Drude formula.

\subsection{Problems of the Langevin equation}
The  Langevin equation (\ref{lan1}) can be integrated and written as:
\beq
u(t) = u(t_0)e^{-\gamma(t-t_0)} + \frac{1}{m}\int_{t_0}^t dt^\p e^{-\gamma (t-t^\p)} R(t^\p).
\label{e3p109}
\eeq
Multiply the above equation with $u(t_{0})$ and shift t to $t_{0}+t$ in the above equation. Then ensemble average becomes
\beq
\la u(t_0)u(t_0 + t)\ra = \la u(t_0)^2\ra e^{-\gamma t}
\eeq
where we have used $ \la u(t_0)\ra \la R(t^\p)\ra = 0$ .  Further taking the time derivative of the above correlation function leads to

\beq
\frac{d}{dt} \la u(t_0) u(t_0+ t)\ra = -\gamma \la u(t_0)^2\ra e^{-\gamma t}.
\eeq
In the limit $t\rta 0$ we have

\beq
\la u(t_0) \dot{u}(t_0)\ra = -\gamma \la u(t_0)^2\ra \neq 0.
\label{statio1}
\eeq
 
Therefore, the time derivative with respect to the intial time$(t_{0})$ of the velocity-velocity correlation function is not zero. It implies that according to the Langevin equation, the velocity correlation function depends on the initial time. But this is contrary to the requirement of stationarity in an equilibrium setting:
\beq
\frac{d}{dt_0} \la u(t_0) u(t_0+ t)\ra =0,
\eeq
i.e.,  the correlation function should not depend on the initial time $t_0$. It only dependent on the difference of time arguments. Simplifying the above equation by differentiating, and then taking the limit $t\rightarrow 0$ we get
\beq
\la u(t_0) \dot{u}(t_0)\ra = 0.
\eeq
Clearly, this is in contradiction to equation (\ref{statio1}). Thus there is serious problem with the Langevin equation. From here it can be concluded that the Drude formula derived from the Langevin equation may give wrong results. It turns out that this conclusion is correct and the Drude scattering rate which is assumed to be constant is not constant in general setting and actually becomes a frequency dependent quantity. The following section points out this problem with the Drude formula.

\section{The generalized Langevin equation and the memory function (time dependent friction coefficient)}
\label{s3p6}
From the previous section we notice that the standard Langevin equation is inconsistent with the stationarity condition. It turns out that this problem can be resolved if the friction coefficient in the Langevin equation is taken as time dependent. In other words constant friction rate ($-m\gamma$) in the standard Langevin equation is to be replaced by time dependent friction coefficient $-m\gamma(t).$ This is equilvalent to considering the colllision history of the particle as will be made clear below. The modified Langevin equation which takes time dependent friction coefficient into account is called the generalised Langevin equation:
\begin{equation}
m \dot{u}(t) = \underbrace{- m \int_{-\infty}^t dt^\p M(t-t^\p) u(t^\p)}_{systematic~part~with~memory} + \underbrace{R(t)}_{random~~part} + \underbrace{E(t)}_{external~drive}.
\label{glan}
\end{equation}

 Here a time dependent friction coefficient $M(t)$ is introduced. We have phenomenologically introduced the generalized Langevin equation(GLE), however, it is well know fact in the non-equilibrium statistical mechanical literature that the GLE can be rigorously derived from equation of motion(EOM) of the brownian particle using projection operator method. We will not review this direction here, and interested readers can consult \cite{zwan}. In the following steps we will derive a modified Drude formula from GLE.


Write GLE in Fourier transform $u(t) = \frac{1}{2\pi}\intmipi d\om u(\om) e^{i \om t}$ domain
\beqar
i\om \frac{1}{2\pi}\intmipi d\om u(\om) e^{i\om t} &=& -\int_{-\infty}^t dt^\p \left(\frac{1}{2\pi}\intmipi M[\om] e^{i\om(t-t^\p)}\right) \left(\frac{1}{2\pi}\intmipi d\om^\p u(\om^\p) e^{i\om^\p t^\p}\right)\nonumber\\
&+& \frac{1}{2\pi}\intmipi d\om (E(\om)+R(\om)) e^{i\om t}, \label{glan1}
\eeqar

where $M[\om]$ is defined as half-Fourier transform (or Fourier-Laplace transform)
\beq
M[\om] = \int_0^\infty dt M(t) e^{-i\om t}.
\label{khjk}
\eeq
Here,  $M(t') =finite$ for $t'\geq0$, and $M(t')=0$ for $t'<0$ (no memory of future). In equation (\ref{glan1}), the upper limit of the integral over $t^\p$ can be extended from $t$ to $+\infty$ as we set $M(t<0) = 0$ and then the integral over $t^\p$ gives a delta function: $\frac{1}{2\pi}\intmipi dt^\p e^{i(\om^\p-\om)t^\p} = \delta(\om^\p -\om)$. This delta function removes the integral over $\om^\p$ and we get the frequency dependent velocity $u(\om)$ in the form : 

\beq
\la u(\om)\ra = \frac{E(\om)}{m (i\om +M[\om])}.
\eeq


In performing the ensemble average we set $\la R(t)\ra = 0,$ if $E(t) = E_0 \cos(\om_0 t)$, then by performing the inverse transformation\footnote{In the algebra one has to use $M[-\om] = M[\om]^\ast$ which can be easily proved from the definition of $M[\om]$.} we get
$\la u(t)\ra = Re \mu(\om_0) E_0 e^{i\om_0 t}.$
Here
\beq
\mu(\om) =\frac{1}{m} \frac{1}{i\om +M[\om]},
\eeq
is called the dynamical mobility. The dynamical conductivity is related to dynamical mobility by $\sigma(\om) = ne^2 \mu(\om)$. Thus 
\beq
\sigma(\om) = \frac{n e^2}{m} \frac{1}{i\om +M[\om]}.
\label{gdf1}
\eeq
{\it This is called the Generalized Drude Formula (GDF) and $M[\om]$ is called the frequency dependent friction coefficient or the memory function.} GDF can also be derived from the Zwanzig-Mori projection operators as shown in following section.
The memeory dependent friction gives Generalised Drude formula. On neglecting this memory we will get back the simple Drude formula. To demonstate it let us consider the memory function to be a delta function in time i.e., $M(t) = \gamma \delta(t +i0)$ (because $M(t)$ is defined only for $t>0$). From equation (\ref{khjk}) we can show that $M[\om]=\gamma$. On substitution it into equation (\ref{gdf1}) we get
\beq
\sigma(\om) = \frac{n e^2}{m} \frac{1}{i\om +\gamma}.
\label{gdf5}
\eeq

This is the same conductivity, which we have obtained in the previous section (equation $(\ref{lan(7)}))$. In a realistic situation $M[\om]$ can be a complicated frequency dependent function, and its computation can be an involved task. In the following section we derive GDF using projection operator technique and then in subsequent sections this technique will be used to derive $M(\om)$ for the important case of metals (using a pertubative procedure given by W\"{o}lfle-G\"{o}tze\cite{gotze}). Important point to remember from this section is that memoryless $M(t)$ leads to simple Drude formula. 

\section{The Zwanzig-Mori projection operator formalism}

In the present section we are going to derive generalised Drude formula using projection operator technique. This provides a powerful computational approach as the MF can be explicitly computed in many situations of interest. In the case of electron scattering in metals we will explicitly show that memory function becomes frequency independent in the low frequency limit and can be identified with the Drude scattering rate. To derive generalised Drude formula using memory function formalism, we need to introduce Mori-Zwanzig projection operators. We start with the Kubo formula \cite{toda,Singh}

\beq
\smnw= V\beta\int_0^\infty  dt e^{i\om t}\la J_{\nu}(0) J_{\mu}(t)\ra.\label{mori2}
\eeq

In the above equation we write $J_{\mu}(t)=e^{i\mcl t}J_{\mu}(0)$ where $\mcl$ is the Liouville operator\footnote{The time evolution of a function of dynamical variables is expressed as $\frac{\pr}{\pr t} f = - \mcl f,$ here $\mcl = i\{H , .\}=\frac{\pr H}{\pr \tbfp}.\frac{\pr}{\pr \tbfq} - \frac{\pr H}{\pr \tbfq}.\frac{\pr}{\pr \tbfp}$. $H$ is the total Hamiltonian of the system. The solution of the above Liouville equation reads $f(t) = e^{i\mcl t}f(0).\label{mori4}$}
Thus
\beq
\smnw = V \beta \int_0^\infty dt e^{i\om t}\la J_\nu(0) e^{i \mcl t} J_\mu(0)\ra.\label{mori5}
\eeq

Time integration can be performed with the result:
\beq
\smnz = V\beta \la J_\nu \frac{i}{z+\mcl} J_\mu \ra.\label{mori5}
\label{3.135}
\eeq
Let us introduce the ``Bra-Ket'' notation of quantum mechanics for convenience, and define the scalar product of two ``vector" A and B as
\beq
\la A | B\ra = \la A^\ast B\ra,\label{mori6}
\eeq 
where we define $\la...\ra = tr (\rho...)$ i.e. the thermodynamical ensemble average. On implementing this notation, equation (\ref{3.135}) can be expressed as 
\beq
\smnz = V\beta \left\la J_\nu \left| \frac{i}{z+\mcl} \right|J_\mu \right\ra.
\label{3.137}
\eeq
Projection operator is defined as
\beq
\mcP \equiv \sum_{\mu^\p} \frac{1}{\la J_{\mu^\p} | J_{\mu^\p} \ra} |J_{\mu^\p} \ra\la J_{\mu^\p} |. \label{3.138}
\eeq
This projection operator projects an arbitrary vector $J_{\mu}$ onto $J_{\mu^\p}$. It is easy to verify that $\mcP\mcP = \mcP$ (the essential property of projection). Introduce another projection operator $\mcQ$ such that $\mcP+\mcQ = \unit$. Rewrite $\mcl$ in equation (\ref{3.137}) as $\mcl = \mcl (\mcP+\mcQ)$
\beq
\smnz = V\beta \left\la J_\nu \left| \frac{i}{\underbrace{z+\mcl \mcQ}_{X} + \underbrace{\mcl \mcP}_{Y}} \right|J_\mu \right\ra.  \label{3.139}
\eeq
With the operator identity
\beq
\frac{1}{X+Y} = \frac{1}{X} - \frac{1}{X} Y \frac{1}{X+Y},  \label{3.140}
\eeq
$\smnz$ can be divided into two terms
\beq
\smnz = i V\beta \underbrace{\left\la J_\nu \left|  \frac{1}{z+\mcl\mcQ} \right| J_\mu \right\ra}_{1st~~term}- i V\beta \underbrace{\left\la J_\nu \left|  \frac{1}{z+\mcl\mcQ} \mcl\mcP \frac{1}{z+\mcl} \right| J_\mu \right\ra}_{2nd~~term}.
\label{3.141}
\eeq
Write the first term as $\frac{1}{z(1+\frac{1}{z} \mcl\mcQ)} = \frac{1}{z} (1-\frac{1}{z} \mcl \mcQ + \frac{1}{z^2} \mcl \mcQ \mcl\mcQ -.......)$.
Except the first term in the expansion, all higher terms vanishes. It is easy to observe this: $\mcQ| J_\mu\ra = \unit |J_\mu \ra - \mcP | J_\mu\ra = \ketjm -\ketjm = 0$. For the second term, insert for the "sandwiched" operator $\mcP$ from equation (\ref{3.138}).  Thus, the equation takes a new form 

\beq
\smnz=i  \beta V\la J_{\nu}\left|\frac{1}{z}\right| J_{\mu}\ra-iV\beta  \chi_0^{-1}\left\la J_\nu \left|  \frac{1}{z+\mcl\mcQ} \mcl \sum_{\mu^\p}\frac{\left| J_{\mu^\p}\left\ra  \right\la  J_{\mu^\p}\right|}{\la J_{\mu^\p} | J_{\mu^\p} \ra} \frac{1}{z+\mcl} \right| J_\mu \right\ra.\label{3.142}
\eeq
Replacing the magnitude form of $\la J_{\nu}|J_{\mu}\ra$ by $\frac{\chi_{0}}{\beta V}\delta_{\nu\mu}$ \cite{Singh} in both terms and manipulating the second term with projection operators, we obtain 
\beq
\smnz=i \frac{\chi_{0}}{z}\unit-V\beta \sum_{\mu^\p}\left\la J_\nu \left| \frac{i}{z+\mcl} \right|J_\mu \right\ra V\beta\chi_{0}^{-1}\frac{1}{z}\left\la J_{\mu^\p}\left|\frac{z}{z+\mcl\mcQ}\mcl \right| J_{\mu}\right\ra.  \label{3.142}
\eeq
We notice that the second term of the above equation can be written as a product of conductivity and the memory function
\beq
\sigma(z)=i \chi_0 \frac{1}{z} \unit -\frac{1}{z} \sigma(z) M(z),    \label{3.143}
\eeq
on simplifying
\beq
\sigma(z)=i \frac{\chi_0}{z + M(z)} = i\frac{\om_p^2}{4\pi}\frac{1}{z+M(z)}, \label{3.144} 
\eeq
where M(z) is the memory function 
\beq
M(z) = V\beta \chi_0^{-1} \left\la J\left| \frac{z}{z+\mcl\mcQ}\mcl \right| J\right\ra.  \label{3.145}
\eeq  
On comparing the above equation (\ref{3.144}) with GDF (equation(\ref{gdf1})) we notice that the memory function can be expressed in terms of projection operator (equation(\ref{3.145})). In the following section we review the G\"{o}tze-W\"{o}lfle perturbation method to evaluate the memory function.


\section{The G\"{o}tze-W\"{o}lfle (GW) Formalism}

Expression given in the above section are formal and cannot be directly used for the computation of the memory function. In 1972, G\"{o}tze and W\"{o}lfle\cite{gotze} presented a perturbative method for the computation of the memory function. They applied their method to the computation of dynamical conductivity of metals. The purpose of this section is to present their method. Before giving the perturbative expansion for memory function we first derive a fundamental equation used by them, i.e.,
\beq
\chi(z) = \chi_0\frac{M(z)}{z+M(z)}.
\label{gw1}
\eeq

Here $\chi(z)$ is the Fourier-Laplace transform of the current-current correlation function

\beq
\chi(z) = i V\int_0^\infty dt e^{izt} \la [J(t), J(0)]\ra.
\label{wg2}
\eeq
Dynamical conductivity $\sigma(z)$ can be written in terms of $\chi(z)$ \cite{Singh}



\beq
\sigma(z) = i\frac{\om_p^2}{4\pi z} - \frac{i}{z} \chi(z).\label{wg4}
\eeq
 We have observed in the previous section that conductivity can be expressed in terms of memory function (equation(\ref{3.144})). An expression for $\chi(z)$ in terms of $M(z)$ can be ontained from equation (\ref{3.144}) and (\ref{wg4}):
\beq
\chi(z) = \frac{\om_p^2}{4\pi}\frac{M(z)}{z+M(z)} =  \chi_0 \frac{M(z)}{z+M(z)} ~~~~~~ or,~~~~ M(z)= \frac{z \chi(z)}{\chi(0)-\chi{(z)}}.
\label{gwmain}
\eeq
A more proof of equation (\ref{gw1}) is given in the Appendix. Memory function can be computed if we compute $\chi(z)$. The above expression is a non-perturbative (exact) expression of the memory function. In the next section we introduce the G\"{o}tze-W\"{o}lfle (GW) perturbative mehtod.

\subsection{The G\"{o}tze-W\"{o}lfle (GW) approximation for the memory function}

For GW metod for the computation of the memory function we need the following EOM \cite{gotze}:
\beqar
z\la\la J;J\ra\ra_z &=& -\la[J,J]\ra - \la\la [J,H];J\ra\ra_z     \nonumber \\
  &= &-\la [J,J] \ra+\la \la J;[J,H]\ra \ra_z
\eeqar

Proof is as follows: 
We have
\beq
z\la\la J;J\ra\ra_z = i V\int_0^\infty dt z e^{izt}\la [J(t),J]\ra = V\int_0^\infty dt \la [J(t),J]\ra \frac{d}{dt}(e^{izt}).
\eeq
On integrating by parts and using $\dot{J}(t) = i[H,J]$, we obtain the first EOM. The second equation can be obtained by using the cyclic property of operators under trace operation. The equal time commutators $[J,J]$ in the above equation are zero. Thus, the first equation can be written as:
\beq
z\chi(z) = -\la\la C; J\ra\ra_z,
\label{gw2}
\eeq
where we define $C= [J, H]$. In the second equation of motion set $J=C$ :
\beq
z\la\la C;J\ra\ra_z = -\la [C, J]\ra +  \la\la C;C\ra\ra_z,
\label{gw3}
\eeq 
or
\beq
\la\la C;J\ra\ra_z  = -   \frac{\la [C, J]\ra -  \la\la C;C\ra\ra_z}{z}.\label{g4}
\eeq
Put $z=0$ in equation (\ref{gw3}) and it leads to $\la [C, J]\ra =  \la\la C;C\ra\ra_0$. Using this and using equation (\ref{gw2}), the equation (\ref{g4}) gives:

\beq
z\chi(z) = \frac{1}{z} (\la\la C;C\ra\ra_0 - \la\la C;C\ra\ra_z).
\eeq
This is an exact expression for the current-current correlation function. Now comes the issue of perturbative expansion. The correlation ($\la\la C;C\ra\ra_z$) has terms proportional to the square of interaction as C is proportional to $H_{interaction}$ $ (C=[J,H]=[J,H_{0}+H_{int}])$. If $H_{int}$ is a small perturbation, then one can linearize the above expression. In equation (\ref{gwmain}) one can expand $M(z)$ as $M(z) = \frac{z\chi(z)}{\chi_0(1-\frac{\chi(z)}{\chi_0})} = \frac{z\chi(z)}{\chi_0}(1+\frac{\chi(z)}{\chi_0}-....)$. Under the condition of small perturbation $||H_{int}||<||H_{0}||$ one can keep the leading order term, and memory function can be approximately expressed as: 

\beq
M(z) \backsimeq \frac{1}{z\chi_0} (\la\la C;C\ra\ra_0 - \la\la C;C\ra\ra_z).
\label{gw100m}
\eeq
This is the central equation used by GW for the computation of dynamical conductivity of a metal in which electron-impurity and electron-phonon scattering is treated as perturbation. In the next two subsections we review their calculation, and then return back to the issue of Drude scattering rate. 

\subsection{Impurity scattering}      

In this subsection we review the GW formalism as applied to the impurity scattering case \cite{gotze}. The case of a simple metal (free electron gas) with impurity scattering is considered. The total Hamiltonian is written as
\beqar
H_{total} &=& H_{electrons} + H_{electron-impurity}\nonumber\\
H_{electrons} &=& \sum_\tbfk \ep(\tbfk) c_\tbfk^\dagger c_\tbfk\nonumber\\
H_{electron-impurity} &=& \frac{1}{N}\sum_{j,\tbfk,\tbfk^\p}\la \tbfk|U^j|\tbfk^\p\ra c_\tbfk^\dagger c_{\tbfk^\p}. \label{im1}
\eeqar
 $N$ is the number of unit cells (we put cell volume to unity). The sum over $j$ is for all the randomly distributed impurities. $U^j$ is the electron scattering potential from $j$th impurity \cite{gotze}. 

Our aim is to calculate memory function (which is generalised Drude scattering rate). We need to compute $C$, and the correlator $\la\la C;C\ra\ra_z$. The operator $C$ is defined as $[J,H_{total}].$ Current density is given by : $J = \sum_\tbfk v_x(\tbfk) c_\tbfk^\dagger c_\tbfk$, where $v_x(\tbfk)$ is the $x-$component of the electron velocity. We consider a case in which the external field is applied along the x-direction and the induced current is also measured in the same direction. As the current operator commutes with the free electron part of Hamiltonian, we get
\beq
C = [J, H_{electron-impurity}] = \frac{1}{N}\sum_{j,\tbfk,\tbfk^\p}\la \tbfk|U^j|\tbfk^\p\ra (v_x(\tbfk)-v_x(\tbfk^\p)) c_\tbfk^\dagger c_{\tbfk^\p}. \label{im2}
\eeq
Define: $ \phi(z) = \la\la C;C\ra\ra_z$ then equation (\ref{gw100m}) can be written as
\beq
M(z) = \frac{1}{z\chi_0} (\phi(0) - \phi(z)).
\label{3.180}
\eeq
$\phi(z)$ will have correlators of the form: written as $\la\la c_\tbfk^\dagger c_{\tbfk^\p} ; c_\tbfp^\dagger c_{\tbfp^\p}\ra\ra$. 

\beq
\la\la c_\tbfk^\dagger c_{\tbfk^\p} ; c_\tbfp^\dagger c_{\tbfp^\p}\ra\ra = i\int_0^\infty dt e^{izt}\la [c_\tbfk^\dagger(t) c_{\tbfk^\p}(t),c_\tbfp^\dg c_{\tbfp^\p}]\ra. \label{im4}
\eeq
Using $c_\tbfk(t) = c_\tbfk e^{-i \ep_\tbfk t}$, the time integration can be performed with the result 
\beq
-\frac{1}{z +\ep_\tbfk -\ep_{\tbfk^\p}}\la [c_\tbfk^\dg c_{\tbfk^\p}, c_\tbfp^\dg c_{\tbfp^\p}]\ra.
\eeq
By using anticommutation relations for $c$ and $c^\dg$, thermal ensemble average can be found \cite{Singh}
\beq
-\frac{1}{z+\ep_\tbfk - \ep_{\tbfk^\p}} (f(\tbfk) - f(\tbfk^\p)).
\eeq 
 $f(\tbfk) =  \la c^\dg_\tbfk c_\tbfk\ra$ is the Fermi-Dirac distribution function. With this simplified expression of correlator, $\phi(z)$ can be written as

\beq
\phi(z) =  \la\la C;C\ra\ra_z =- 2\frac{1}{N^2}\sum_{i,j,\tbfk,\tbfk^\p}\la \tbfk|U^i_1|\tbfk^\p\ra \la \tbfk^\p|U^j_1|\tbfk\ra   (v_x(\tbfk)-v_x(\tbfk^\p))^2 \frac{f(\tbfk) -f(\tbfk^\p)}{z+\ep_\tbfk -\ep_{\tbfk^\p}}.
\eeq
The summations over (i,j) run over the total number of impurities ($N_{imp}$). For $i\ne j,~~\phi(z)$ is proportional to $(\frac{N_{imp}}{N})^2$. If the impurity concentration $c=\frac{N_{imp}}{N}$ is very small, then this term can be neglected as compared to the term proportional to $c$ (the diagonal elements contribution). Thus, in the leading order:
\beq
\phi(z) = -2\frac{c}{N}\sum_{\tbfk,\tbfk^\p} |\la \tbfk|U_1|\tbfk^\p\ra|^2 \{v_x(\tbfk)-v_x(\tbfk^\p)\}^2 \frac{f(\tbfk) -f(\tbfk^\p)}{z+\ep_\tbfk -\ep_{\tbfk^\p}}.\label{im7}
\eeq
 
Here the summation over the impurity sites $i$ is also performed because there are no correlation effects considered \cite{gotze,Singh}. And the problem reduces to one impurity problem. Further simplification can be done using isotropy in the free electron case 
${\bf v}^2 = 3 v_x^2$ and writing ${\bf v} = \tbfk/m$. Substituting equation (\ref{im7}) into equation (\ref{3.180}) we get 

\beq
M(z) = \frac{2}{3 z\chi_0}\frac{c}{m^2 N}\sum_{\tbfk,\tbfk^\p} |\la \tbfk|U_1|\tbfk^\p\ra|^2 \{\tbfk-\tbfk^\p\}^2 (f(\tbfk) -f(\tbfk^\p))\left(\frac{1}{z+\ep_\tbfk -\ep_{\tbfk^\p}} - \frac{1}{\ep_\tbfk -\ep_{\tbfk^\p}}\right). \label{im8}
\eeq
Substitute $z=\om+i\varepsilon$ and perform the limit $\varepsilon\rta0$, the above expression reduces to\footnote{Use the identity $\lim_{\varepsilon\rta0}\frac{1}{a\pm i\varepsilon} = \frac{1}{a}\mp i\pi \delta(a)$.} 

\beq
Im M(\om) = \frac{2}{3} \pi (\frac{N_{imp}}{N_e}) \frac{1}{m N^2} \sum_{\tbfk,\tbfk^\p} |U(\tbfk,\tbfk^\p)|^2 (\tbfk-\tbfk^\p)^2 (\frac{f(\tbfk)-f(\tbfk^\p)}{\om})\delta(\om +\ep_\tbfk -\ep_{\tbfk^\p}). \label{im9}
\eeq

Using $\frac{1}{N}\sum_{\tbfk} \longrightarrow \int\frac{d^3k}{(2\pi)^3},$ equation (\ref{im9}) becomes

\beqar
Im M(\om) &=& \frac{2}{3}\pi (\frac{N_{imp}}{N_e})\frac{1}{m} \int\frac{d^3k}{(2\pi)^3} \int\frac{d^3k^\p}{(2\pi)^3} |U(\tbfk,\tbfk^\p,\theta)|^2 (k^2 +{k^\p}^2 - 2 k k^\p \cos(\theta))\nonumber\\
&\times& \left(\frac{f(\tbfk)-f(\tbfk^\p)}{\om}\right)\delta(\om + \ep_\tbfk -\ep_{\tbfk^\p}). \label{im10}
\eeqar

By changing variables $\ep = \frac{k^2}{2 m}~~(\hbar=1)$ and performing the integration over $\ep^\p=\frac{{k^\p}^2}{2 m}$ and using the properties of the delta function we obtain:
\beqar
Im M(\om) &=&\frac{2}{3}\pi (\frac{N_{imp}}{Ne})\frac{(2m)^3}{(2\pi)^4}\int_0^\infty d\ep \sqrt{\ep(\ep+\om)}\int_0^\pi d\theta \sin\theta |U(\ep,\ep+\om,\theta)|^2\nonumber\\
 &\times&(2\ep+\om-2\sqrt{\ep(\ep+\om)}\cos\theta) \left(\frac{f(\ep)-f(\ep+\om)}{\om}\right). \label{im11}
\eeqar

This is a finite frequency result. In the limit $\om\rta 0$, the last factor gives a delta function and the integral over energy can be performed with the result

\beq
 Im M(0) =\frac{4}{3}\pi (\frac{N_{imp}}{Ne})\frac{(2m)^3}{(2\pi)^4} \ep_F^2 \int_0^\pi d\theta \sin\theta |U(\ep_F,\theta)|^2 (1-\cos\theta). \label{im12}
\eeq 
With the angle dependence of the scattering potential taken into account it is clear that the scattering rate ($\frac{1}{\tau} = Im M(0)$) has very small contribution from small angle scattering (when $\theta$ is small). Thus, small angle scattering is not efficient in the momentum degradation. The very same conclusion is obtained when one deal with the solution of the Boltzmann equation beyond Relaxation Time Approximation (RTA) in an isotropic medium \cite{Singh,ziman,madelung,wooten}. Thus the above expression from memory function formalism goes beyond the RTA \cite{Singh}.

Next, consider that the scattering potential is independent of the angle between $\tbfk$ and $\tbfk^\p$. The above equation (\ref{im12}) can be further simplified:

\beq
Im M(0) = \frac{2}{3}\pi \frac{N_{imp}}{N_e} (U \rho_F)^2 \ep_F.
\eeq
Thus the real part of conductivity from equation (\ref{3.144}) is given by

\beq
Re\sigma(\om) = \frac{\om^2_p}{4\pi}\frac{Im M(0)}{\om^2 + (Im M(0))^2}. \label{im16}
\eeq
This is nothing but the Drude formula. We identify that 

\beq
\frac{1}{\tau}=Im M(0) = \frac{1}{\taud} = \frac{2}{3}\pi \frac{N_{imp}}{N_e} (U \rho_F)^2 \ep_F.
\label{memdrude}
\eeq

Thus, the Drude scattering rate increases linearly (in the leading order) with impurity concentration and with the square power of the electron-impurity scattering potential. {\it The most important point is that we recover the Drude formula in the low frequency limit. The low frequency here means $\om<<\ep_F$ (the Fermi energy). If Fermi energy is in $eV$s then the valid frequency regime of simple Drude formula is below say infrared frequencies. Thus} 

\beq
\frac{1}{\tau} {\rm~~ is~frequency~independent,~when~~} \hbar\om<<\ep_F.
\eeq
This defines the regime of applicability of the Drude formula in metals \cite{wooten}.

\subsection{Phonon scattering}

Next application of the WG formalism is the computation of the Memory Funtion (MF) in the case of electron-phonon scattering in metals. The leading cause of electrical resistance at ambient temperatures is due to the electron-phonon scattering. At ambient temperature electrical resitivity is linearly proportional to tempeature i.e., $\rho \propto T.$ In a pure metalic sample at lower temperature ($T<<\Theta_{D}$, Debye Temperature), resitivity is proportional to fifth power of temperature ( $\rho \propto T^5)$. These temperature dependencies have been experimentally verified. Theoretically these can be obtained from the solution of the Bloch-Boltzmann equation \cite{Singh,ziman,madelung}. The Bloch-Boltzmann equation formulates the electron-phonon scattering in a semiclassical way. We will notice in this section that these temperature dependencies can also be obtained from memory function formalism. Before we start the calculation of the MF in the present case of electron-phonon scattering, let us note down its basic assumptions. For conduction electrons, a free electron gas is assumed. The phonon considered are long wavelength longitudinal acoustic phonons. For electron-phonon interaction the Fr\"{o}hlich Hamiltonian is used:

\beqar
H_0 &=& H_{el} + H_{ph},\nonumber\\
H_{el-ph} &=& \sum_{\tbfk,\tbfkp} (D(\tbfk-\tbfkp) c_\tbfk^\dg c_\tbfkp b_{\tbfk-\tbfkp} + h.c.).
\eeqar
Here $H_0$ is the free electron and free phonon part of the Hamiltonian with $H_{el} =\sum_\tbfk c_\tbfk^\dg c_\tbfk $ and $H_{ph} = \sum_\tbfq \om_\tbfq (b_\tbfq^\dg b_\tbfq + 1/2)$. The second equation represents the electron phonon interaction. The coefficient $D(\tbfk-\tbfkp)$ for acoustic phonon interaction is given by 

\beq
D(\tbfq) = \frac{1}{\sqrt{2 m_{ion} N \om_\tbfq}} q C_q,
\eeq
$N$ is the total number of unit cells, $m_{ion}$ is the ionic mass,  and $\om_\tbfq$ is the phonon frequency. $C_q$ is the electron-phonon coupling constant and it is a slowly vaying function of $q$ \cite{madelung}. As before we work with units $\hbar=k_{B}=1$. To compute the memory function we need to compute C:
\beq
C=[J_x, H_{el-ph}] = \sum_{\tbfk,\tbfkp,\tbfpp} v_x(\tbfk) D(\tbfp-\tbfpp)[c_\tbfk^\dg c_\tbfk, c_\tbfp^\dg c_\tbfpp b_{\tbfp-\tbfpp}]- h.c.
\eeq
 Using the anticommutation properties of $c_\tbfk$, and $c_\tbfk^\dg$, and using the identity $[AB,C] = A[B,C] + [A,C]B$ we get

\beq
C = \sum_{\tbfp,\tbfpp} (v_x(\tbfp)-v_x(\tbfpp)) D(\tbfp-\tbfpp)c_\tbfp^\dg c_\tbfpp b_{\tbfp-\tbfpp} - h.c.
\eeq
Our next step is to calculate $\phi(z) = \la\la C;C\ra\ra_z$. Out of four terms in it only non-diagonal terms survive:
\beqar
\phi(z) &=&\nonumber\\
& -&\sum_{\tbfk,\tbfkp,\tbfp,\tbfpp} (v_x(\tbfk)-v_x(\tbfkp))(v_x(\tbfp)-v_x(\tbfpp)) D(\tbfk-\tbfkp) D^\ast(\tbfp-\tbfpp) \la\la c_\tbfk^\dg c_\tbfkp b_{\tbfk-\tbfkp};b^\dg_{\tbfp-\tbfpp}c^\dg_\tbfpp c_\tbfp\ra\ra\nonumber\\
&-&\sum_{\tbfk,\tbfkp,\tbfp,\tbfpp} (v_x(\tbfk)-v_x(\tbfkp))(v_x(\tbfp)-v_x(\tbfpp)) D^\ast(\tbfk-\tbfkp) D(\tbfp-\tbfpp) \la\la c_\tbfkp^\dg c_\tbfk b^\dg_{\tbfk-\tbfkp};b_{\tbfp-\tbfpp}c^\dg_\tbfp c_\tbfpp\ra\ra.\nonumber\\
\eeqar

 The correlation function in its expanded notation is given by:

\beq
\la\la c_\tbfk^\dg c_\tbfkp b_{\tbfk-\tbfkp};b^\dg_{\tbfp-\tbfpp}c^\dg_\tbfpp c_\tbfp\ra\ra = i\int_0^\infty dt e^{i z t}\la [c_\tbfk^\dg(t) c_\tbfkp(t^\p) b_{\tbfk-\tbfkp}(t), b^\dg_{\tbfp-\tbfpp}c^\dg_\tbfpp c_\tbfp]\ra.
\eeq
Using $c_\tbfk(t) = c_\tbfk e^{-i\ep_\tbfk t}$, the time integration and the ensemble average can be performed. The result is

\beqar
\phi(z) &=& \frac{2}{3}\frac{1}{m^2}\sum_{\tbfk,\tbfkp} |D(\tbfk-\tbfkp)|^2 |\tbfk-\tbfkp|^2\{f(1-f^\p)(1+n_{-})-(1-f)f^\p n_{-}\}\nonumber\\
&\times&\left(\frac{1}{\ep -\ep^\p -\om_{-} +z} +\frac{1}{\ep -\ep^\p -\om_{-} -z}\right).
\eeqar
Here a factor of 2 in the coefficient comes from the spin summations. We have used short hand notation: $f$ means $f(\tbfk)$; $f^\p$ means $f(\tbfkp)$; $n_{-}$ is for $n_{\om_{\tbfk-\tbfkp}}$, and $\om_{-}$ is for $\om_{\tbfk-\tbfkp}$. On inserting the above equation into (\ref{3.180}) and on simplifying we get:

\beqar
Im M(\om) &=& \frac{2}{3}\pi\frac{1}{m^2\chi_0} \sum_{\tbfk,\tbfk^\p} |D(\tbfk-\tbfk)|^2 |\tbfk-\tbfkp|^2 (1-f)f^\p n_{-}\nonumber\\
&\times&\left(\frac{(e^{\beta \om}-1)}{\om} \delta(\ep-\ep^\p -\om_{-} +\om) - ...({\rm terms~with~}\om \rta -\om)...\right).
\eeqar


Convert momentum sums into integrals. To simplify the momentum integrals, change $|\tbfk-\tbfkp|$ into q integral, by introducing $\int dq \delta(q-|\tbfk-\tbfkp|)$ and writing integration variables k and $k^\p$ as $\sqrt{2 m \ep}$ and $\sqrt{2 m \ep'}$, we obtain



\beqar
&&Im M(\om) =\nonumber\\
&&\frac{2}{3}\pi\frac{N m^2}{ (2\pi)^3  m_{ion} N_e}\int_0^\infty dq \frac{C_q^2}{\om_q}q^4\int_0^\infty d\ep \sqrt{\ep}\int_0^\infty d\ep^\p \sqrt{\ep^\p}\int_0^\pi d\theta \sin\theta \delta\left(q-\sqrt{2m}\sqrt{\ep+\ep^\p-2\sqrt{\ep\ep^\p}\cos\theta}\right)\nonumber\\ 
&\times&(1-f) f^\p n_{-}\left(\frac{(e^{\beta \om}-1)}{\om} \delta(\ep-\ep^\p -\om_{-} +\om) - ...({\rm terms~with~}\om \rta -\om)...\right). \label{ph6}
\eeqar

The integral over $\theta$ can be simplified in the following way. The presence of the Fermi factors $(1-f)f^\p$ cause the integrand to have a finite value only in a zone around the Fermi surface of a strip of width $2 k_B T.$ And outside of this zone integrand is vanishingly small. Thus $\ep$ and $\ep^\p$ can be approximately replaced by $\ep_F$. On implimenting all these steps, the $\theta$ integral reduces to:  
\beq
\int_0^\pi d\theta \sin\theta \delta(q - \sqrt{2}k_F \sqrt{1-\cos\theta}). \label{ph7}
\eeq
 This elementary integration can be performed with the result $\frac{q}{k_F^2}$ (notice that $0<q<k_F$). Insert the value $\frac{q}{k_F^2}$ of the $\theta$-integral into equation (\ref{ph6}). Perform the $\ep^\p$ integral using the property of the delta functions. Then, perform the integrations over $\ep$ using elementary methods\footnote{use $\intmipi dx \frac{e^x}{e^x+1} \frac{1}{e^{x+a}+1} = \frac{a}{e^a-1}$.}. Finally, we obtain

\beqar
Im M(\om) &=& \frac{1}{8}\pi^3N (m m_{ion} k_F^5)^{-1}\rho_F^2 \int_0^{q_D} dq q^5 \frac{C_q^2}{\om_q} \left(\frac{1}{e^{\beta \om_q}-1}\right)\nonumber\\
&\times& \left\{ \frac{(1-\om_q/\om)(e^{\beta\om}-1)}{e^{\beta(\om-\om_q)}-1} + ...({\rm terms~with~}\om \rta -\om)... \right\}.
\label{ph8}
\eeqar
The above equation (\ref{ph8}) is GW's equation (\ref{im4}) in \cite{gotze}. For further simplifications we assume the linear approximation to phonon spectrum $\om_q = c_s q$ (i.e., the Debye approximation) and $C_q$ is assumed constant $C_q=\frac{1}{\rho_F}$ \cite{gotze}. In the following subsections we analyse the above expression in various special cases of interest. We will notice that Bloch-Boltzmann equation results can also be obtained by this metod. Let us first consider the D.C. case.

 \subsubsection{D.C. case}

In the $\om\rta0$ limit, the expression in the curly brackets in the above equation (\ref{ph8}) reduces to $2\beta\frac{\om_q}{1-e^{-\beta\om_q}}.$ Thus
\beq
Im M(0) = \frac{1}{4}\pi^3 N (m m_{ion} k_F^5)^{-1}\int_0^{q_D} dq \frac{q^5 e^{\beta \om_q}}{(e^{\beta\om_q}-1)^2}.
\eeq
To simplify, write $x = \beta \om_q$, and  $\beta \om_D = \Theta_D = \frac{\Theta_D}{T}$:
\beq
Im M(0) = \frac{1}{4}\pi^3 N q_D^6 (m m_{ion} k_F^5\Theta_D)^{-1} (T/\Theta_D)^5 J_5(\Theta_D/T),~~{\rm~where~~~}J_5(y) = \int_0^y dx \frac{x^5 e^x}{(e^x-1)^2}. 
\eeq
The above expression is the DC scattering rate (equilvalent to the Drude scattering rate). It can be further analyzed into two special cases. In the high temperature limit, $T>>\Theta_D$, we have $J_5(y)\simeq \frac{1}{4} y^4$, thus
\beq
Im M(0) = \frac{1}{\tau}= \frac{1}{16} \pi^3 N q_D^6 (m m_{ion} k_F^5 \Theta_D)^{-1} (T/\Theta_D) \propto T.
\eeq
So, we observe that in high temperature approximation (when the temperature (T) is much greater than the Debye temperature $\Theta_{D}))$, the scattering rate or, the imaginary part of the memory function has $T-$ linear temperature dependence. This agrees with the result as obtained from the Bloch-Boltzmann equation in the high temperature limit \cite{Singh,ziman}. In the low temperature limit, i.e, $T<<\Theta_D$, we have $J_5(y)\simeq 124.4$, and the MF is
\beq
Im M(0) = \frac{1}{\tau}= 31.1 \pi^3 N q_D^6 (m m_{ion} k_F^5 \Theta_D)^{-1} (T/\Theta_D)^5 \propto T^5.
\eeq
This again agrees with the famous Bloch-Boltzmann's $T^5$ law of phonon scattering. Thus all the known results in the DC limit are reproduced (Table 1:).
\begin{table}[h]
\caption{Temperature dependence of memory function in DC limit.} \label{tab1}
\begin{center}
\begin{tabular}{|c||c|c|c|}
\hline
\multirow{3}{*}{Temperature limit}
 & \multicolumn{2}{c||}{D.C. case} \\      \cline{2-3}
 & $ J_{5}(y) $ & $ImM(0)$ \\  \hline
 $T>>\Theta_{D}$ &$ \simeq \frac{y^4}{4}$ & $\propto T$  \\  \hline
 $T<<\Theta_{D}$ &$ \simeq 124.4$ & $\propto T^5$ \\   \hline
\end{tabular}
\end{center}
\end{table}
\subsubsection{A.C. case}
The integral in equation (\ref{ph8}) in $\om \neq 0$ limit is difficult to perform (although it can be expressed in terms of PloyLog functions). We consider a simpler and relevant case of high frequencies $\om >>\Theta_D$, and $\om>>T$. Then the expression in curly brackets in equation (\ref{ph8}) can be simplified to give $e^{\beta\om_q} +1$. On simplifying the integral over q we obtain
\beq
Im M(\om) \simeq \frac{1}{4}\pi^3 N q_D^6 (m m_{ion} k_F^5)^{-1} (T/\Theta_D)^5 J(\Theta_D/T),~~~J(y) =\frac{1}{2}\int_0^y dx x^4 \coth(x/2). \label{ac1}
\eeq
In the low temperature limit, i.e. $T<<\Theta_D$, we have $J(y) \simeq \frac{1}{10} y^5$, thus
\beq
Im M(\om) \simeq \frac{1}{40}\pi^3 N q_D^6 (m m_{ion} k_F^5)^{-1}. \label{ac2}
\eeq 
We observe that at high frequency $\om>>\Theta_{D}$, but at lower temperature $T<<\Theta_{D}$, the scattering rate reduces to a constant value independent of frequency and temperature (Table 2:). And in the high temperature limit i.e., $T>>\Theta_D,$ we have $J(y)\simeq\frac{1}{4} y^4$. The temperature dependence of MF is given by
\beq
Im M(\om) \simeq \frac{1}{16}\pi^3 N q_D^6 (m m_{ion} k_F^5)^{-1} (T/\Theta_D) \propto T. \label{ac3}
\eeq
This behaviour is depicted in figure 2. An important point to be noticed is that even at zero temperature scattering rate is not zero. This non-zero scattering rate is due to an important mechanism called the Holstein mechanism.

In the limit $T\rta0$, equation (\ref{ph8}) can be simplified for all values of $\om$. In this limit, the expression in the curly bracket in (\ref{ph8}) gives $(1-\om_q/\om)e^{\beta\om_q}$ if $\om>\om_q,$ and $(\om_q/\om-1)e^{\beta\om}$ if $\om<\om_q$. After performing the q integral we get:

\beq
Im M(\om) \simeq \frac{1}{240}\pi^3 N q_D^6 (m m_{ion} k_F^5 \Theta_D)^{-1}\times\left\{ \begin{array}{lr}(\om/\Theta_D)^5& : |\om| <\Theta_D\\
6-5 (\om_D/\om) & : |\om| >\Theta_D. \label{ac4}
\end{array}
\right. 
\eeq
 In the limit $T\rta0$, the non-vanishing value of the generalised Drude scattering rate (or memory function) has a very important physical meaning as mentioned before. In a pure sample at zero temperature there are no thermally excited lattice vibrations or phonons, thus one would naviely expect that electron scattering would be prohibited and the scattering rate would vanish. But it is shown for the first time by Ted Holstein that there is an important mechanism of momentum randomization even at zero temperature \cite{ted}. The mechanism involves the simultaneous absorption of a photon and creation of an electron-hole pair along with an acoustic phonon. Thus the mechanism---now called the Holstein mechanism---involves the creation of a phonon and it occurs at finite frequencies. From equation (\ref{ac4}) we observe that G\"{o}tze-W\"{o}lfle formalism is capable of caputuring this physical effect. At low frequencies $|\om|<\Theta_{D}$, the scattering rate grows as $\om^5$ and then it saturates when $\om>>\Theta_{D}$. The transition happens around $\Theta_{D}$ (Table2). The generalised Drude scattering rate can be obtained from the experimental data of reflectance using the memory function formalism. If experimentally obtained scattering rate shows transition behaviour around $\Theta_{D}$, then it can be derived that the dominating scattering mechanism is due to phonons. This perticular issue is very important in the field of strange metals such as observed in normal states of cuprate high temperature superconductors. Absence of $\Theta_{D}$ points to other scattering mechanisms.

\begin{figure}[h!]
\begin{center}
\includegraphics[height = 7cm, width =10cm]{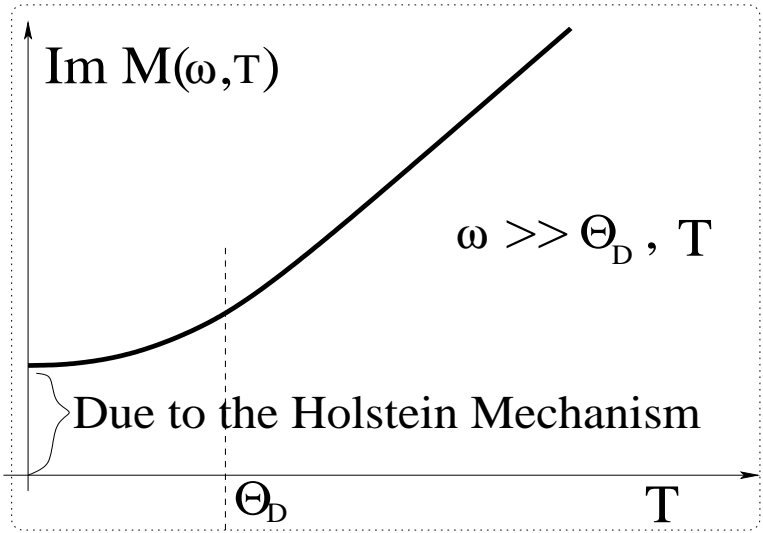}
\caption{A depiction of the Holstein mechanism}
\label{fig1}
\end{center}
\end{figure}
\begin{table}[h]
\caption{Temperature dependence of the memory function in the AC limit} 
\begin{center}
\begin{tabular}{|c||c|c|c|}
\hline
\multirow{3}{*}{($\om>>\Theta_{D}$)\newline Temperature limit}
 & \multicolumn{2}{c||}{A.C. case} \\      \cline{2-3}
 & $ J_{5}(y) $ & $ImM(0)$ \\  \hline
 $T>>\Theta_{D}$ &$ \simeq \frac{y^4}{4}$ & $\propto T$  \\  \hline
 $T<<\Theta_{D}$ &$ \simeq \frac{y^5}{10}$ & $independent ~of~ T$ \\   \hline
\end{tabular}
\end{center}
\end{table}
We also notice that the generalised Drude scattering rate (equation (\ref{ac4})) strongly deviates from the simple Drude scattering rate (which is just a constant parameter). From equation 
(\ref{im16}) the real part of conductivity is given by:

\beq
Re\sigma(\om) = \frac{\om^2_p}{4\pi}\frac{Im M(\om)}{\om^2 + (Im M(\om))^2}. \label{ac5}
\eeq
Where $ImM(\om)$ is given by the expression (\ref{ac4}). If the sample also has dilute impurities, then in the linear order $ImM_{total}(\om)=Im M_{imp}(\om)+ImM(\om)$, which in the 
$\om\rta 0$ limit goes back to Drude formula (equation (\ref{mom2})).
\section{Summary of MF formalism}
In this brief overview of the memory function formalism we sketched the derivations of the generalised Drude formula by two methods. In the first method it is derived from the generalised Langevin equation. An advantage of this route is that we are able to appreciate the problems of the standard Drude formula and the corresponding standard Langevin equation. Then in the second route we reviewed the Zwanzig-Mori projection operator method, and showed that dynamical conductivity from the Kubo formula can be written in a form resembling the generalised Drude formula. In doing that the memory function is expressed through projection operators. We then reviewed a very useful method for the computation of the memory function put forward by G\"{o}tze-W\"{o}lfle (GW). Two applications of this method in the computation of dynamical conductivity of metals are reviewed. In the first application of the electron-impurity scattering we notice that the Drude scattering rate is frequency dependent. It can be taken as frequency independent only when the system is probed with frequencies much less than $\frac{\ep_{F}}{\hbar}.$ We also noticed that the DC scattering rate includes $1-\cos\theta$ factor which stresses the importance of large angle scattering in momentum degradation. This factor appears in the solution of the Bloch-Boltzmann equation when vortex corrections are taken into account. In the elecron-phonon scattering case we notice that in the DC limit the memory function formalism reproduces all the known results of the Bloch-Boltzmann equation ($\rho\propto T~when ~T>>\Theta_{D}, ~and~ \rho\propto T^5 ~when~T<<\Theta_{D}).$ In the AC limit it also reproduces the Holstein mechanism of electromagnetic energy absorption at zero temperature.

\section{Appendix: The G\"{o}tze-W\"{o}lfle (GW) Formalism}
We need to show that
\beq
\chi(z) = \chi_0\frac{M(z)}{z+M(z)},
\label{wg1}
\eeq

where $\chi(z)$ is 
\beq
\chi(z) = i V\int_0^\infty dt e^{izt} \la [J(t), J(0)]\ra.
\label{wg2}
\eeq
Let us define

\beq
\zeta(t) = i\la [J(t),J(0)]\ra.
\eeq 
Using the cyclic property of trace $\zeta(t)$ can be written as
\beq
i tr(\rho J(t) J) - i tr(\rho J J(t)) = i tr(J(t) [J,\rho]).
\label{wg3}
\eeq

Use Kubo's Identity $e^{\beta H} [J, e^{-\beta H}] = \int_0^\beta e^{\lambda H} [H, J]e^{-\lambda H},$ and write $[J,\rho] $ it in the following way

\beq
[J,\rho] = \int_0^\beta d\lambda \rho [H, J(-i\lambda)].
\eeq
 Using $\dot{A} = i [H, A]$, the above commutator takes the form 

\beq
[J, \rho] = -i \int_0^\beta d\lambda \rho \dot{J}(-i\lambda).
\eeq

Inserting this expression in equation (\ref{wg3}), $\zeta(t)$ takes the form

\beq
\zeta(t)  = \beta \la\dot{J}(-i\lambda) J(t)\ra^\diamond.
\eeq 
Here we have introduced a notation: $\la...\ra^\diamond = \frac{1}{\beta}\int_0^\beta d\lambda tr (\rho...)$. With this, the equation (\ref{wg2}) can be written as

\beq
\chi(z) = V \int_0^\infty dt e^{izt}\zeta(t) =  V\beta \int_0^\infty dt e^{izt} \la \dot{J}(-i\lambda) J(t)\ra^\diamond.
\label{wg10}
\eeq

For the next part of the proof, we will use an alternative representation of conductivity from Kubo's formula \cite{kubo}: 

\beq
\sigma(z) = -\frac{\om_{p}^2}{4 \pi iz} - \frac{\beta V}{iz} \int_0^\infty dt e^{izt} \la J(-i\lambda)\dot{J}(t)\ra^\diamond.
\label{wg11}
\eeq

We want to find a connection between the last two equations ( (\ref{wg10}) and (\ref{wg11})). Only problem is that the integrands differ w.r.t. time differentiation. To sort this out, we use the stationarity property of the current-current correlation function, i.e., $\la J(t_0)J(t_0+t)\ra$ must be independent from the initial time $t_0$ or
\beq
\frac{d}{dt_0}\la J(t_0) J(t_0 +t + i\lambda)\ra^\diamond = 0 = \la \dot{J}(t_0) J(t_0 +t + i\lambda)\ra^\diamond + \la J(t_0) \dot{J}(t_0 +t + i\lambda)\ra^\diamond. \label{wg12}
\eeq

Thus at $t_0=0$
\beq
\la \dot{J}(0) J(t + i\lambda)\ra^\diamond = - \la J(0) \dot{J}(t + i\lambda)\ra^\diamond.
\eeq
Using this, from equation (\ref{wg10}) and (\ref{wg11}) we obtain the required result.

\end{document}